\documentclass[twocolumn,superscriptaddress,showpacs,oneside,prl]{revtex4}
\usepackage{amsmath}
\usepackage{amssymb}
\usepackage{amsfonts}
\usepackage{graphicx}

\begin{document}

\title{Experimental implementation of high-fidelity unconventional geometric
quantum gates using NMR interferometer}
\author{Jiangfeng Du}
\email{djf@ustc.edu.cn} \affiliation{Hefei National Laboratory for
Physical Sciences at Microscale \& Department of Modern Physics,
University of Science and Technology of China, Hefei, Anhui
230026, China} \affiliation{Department of Physics, National
University of Singapore, 2 Science Drive 3, Singapore 117542}
\author{Ping Zou}
\affiliation{Hefei National Laboratory for Physical Sciences at
Microscale \& Department of Modern Physics, University of Science
and Technology of China, Hefei, Anhui 230026,  China}
\author{Z. D. Wang}
\email{zwang@hkucc.hku.hk}
\affiliation{Department of Physics \& Center of Theoretical and Computational Physics,
University of Hong Kong, Pokfulam Road, Hong Kong, China}
\date{\today}

\begin{abstract}
Following a key idea of unconventional geometric quantum
computation developed earlier  [Phys. Rev. Lett. 91, 197902
(2003)],  here we propose a more general
scheme in such an intriguing way: $\gamma _{d}=\alpha _{g}+\eta \gamma _{g}$%
, where $\gamma _{d}$ and $\gamma _{g}$ are respectively the dynamic and
geometric phases accumulated in the quantum gate operation, with $\eta$ as a
constant and $\alpha _{g}$ being dependent only on the geometric feature of
the operation. More arrestingly, 
we demonstrate the first experiment to implement a universal set of such
kind of generalized unconventional geometric quantum gates with high
fidelity in an NMR system.
\end{abstract}

 \maketitle



Quantum computation has been paid intensive interest for the past decade
because quantum computers are believed to be much more powerful and
efficient than their classical counterparts due to their quantum nature\cite%
{Shor1999}. Significant progresses have recently been achieved in the field
of quantum computing. Nevertheless, there are still great challenges in
physical implementation of quantum computation. One of them is to suppress
the noises in quantum gates to an acceptable level, which is essential to
build a scalable quantum computer. Recently, a promising approach based on
geometric phases\cite{Berry1984,Aharonov,Zhu2000} was proposed to achieve
built-in fault-tolerant quantum gates with higher fidelities\cite%
{Zanardi,Jones,X.B.Wang,Falci,Zhu,Cirac,S.L.Zhu} since the geometric phase
depends only on the global feature of the evolution path and is believed to
be robust against local fluctuations. On the other hand, in the same spirit,
an interesting unconventional geometric quantum computation(GQC) scheme was
proposed\cite{ZhuWang03}; such kind of two-qubit gate was indeed reported
experimentally in trapped ions\cite{Leibfried} and was designed with
superconducting qubits\cite{ZhuWang05}. In this scheme, the dynamic phase $%
\gamma _{d}$ is ensured to be proportional to the geometric phase $\gamma
_{g}$, namely, $\label{proportional}\gamma _{d}=\eta \gamma _{g},\ \ (\eta
\neq 0,\ -1)$ with $\eta $ as a proportional constant.


In this paper, we propose that the above unconventional GQC scheme\cite%
{ZhuWang03} can be further generalized in such an intriguing manner: $%
\gamma_d=\alpha_g + \eta\gamma_g$, where $\alpha_g$ is a coefficient
dependent only on the geometric feature of the quantum evolution path in the
gate operation. It is elaborated that this generalized unconventional GQC
can be realized in physical systems like NMR. In particular, we report the
first experimental implementation of a universal set of such kind of
unconventional geometric gates with high fidelity in an NMR system.

Before we present our new results, let us first elucidate how to realize a
single-qubit gate 
with the generalized unconventional geometric phase shift in the cyclic
evolution\cite{S.L.Zhu}. For an one-qubit system, consider two orthogonal
cyclic states $\left |\psi_{+}\right\rangle$ and $\left|\psi_{-}\right%
\rangle $, which satisfy the relation $U(\tau)\left|\psi_{\pm}\right%
\rangle=exp(\pm i\gamma)\left|\psi_{\pm}\right\rangle$, where $\gamma$ is
the total phase accumulated and $U(\tau)$ is the evolution operator of a
cyclic evolution with $\tau$ as the periodicity. We can write $%
\left|\psi_{+}\right\rangle=e^{-i\frac{\phi}{2}}\cos\frac{\chi}{2}%
\left|\uparrow\right\rangle+e^{i\frac{\phi}{2}}\sin\frac{\chi}{2}%
\left|\downarrow\right\rangle$ and $\left|\psi_{-}\right\rangle=-e^{-i\frac{%
\phi}{2}}\sin\frac{\chi}{2}\left
|\uparrow\right\rangle+e^{i\frac{\phi}{2}%
}\cos\frac{\chi}{2}\left |\downarrow\right\rangle$ , where ($\chi$, $\phi$)
are the spherical coordinates of the state vector on the Bloch sphere
(Fig.1), $\left |\uparrow\right\rangle$ and $\left
|\downarrow\right\rangle$
are the two eigenstates of the $z$-component of the spin-1/2 operator ($%
\sigma_{z}/2$) and they constitute the computational basis for the qubit.
For an arbitrary input state denoted as $\left
|\psi_{in}\right%
\rangle=a_{+}\left |\psi_{+}\right\rangle+a_{-}\left
|\psi_{-}\right\rangle$
with $a _{\pm}=\langle\psi_{\pm}\left
|\psi_{in}\right\rangle$, after the
cyclic evolution for the $\left |\psi_{+}\right\rangle$ ($%
\left
|\psi_{-}\right\rangle$) state, the output state is $\left
|\psi_{out}\right\rangle=U(\gamma,\chi,\phi)\left
|\psi_{in}\right\rangle$,
where

\begin{equation}
U=\left(%
\begin{array}{cc}
e^{i\gamma}\cos^{2}\frac{\chi}{2}+e^{-i\gamma}\sin^{2}\frac{\chi}{2} &
ie^{-i\phi}\sin\gamma\sin\chi \\
ie^{i\phi}\sin\gamma\sin\chi & e^{i\gamma}\sin^{2}\frac{\chi}{2}%
+e^{-i\gamma}\cos^{2}\frac{\chi}{2} \
\end{array}%
\right).
\end{equation}

\begin{figure}[b]
\begin{center}
\includegraphics[width=0.8\columnwidth]{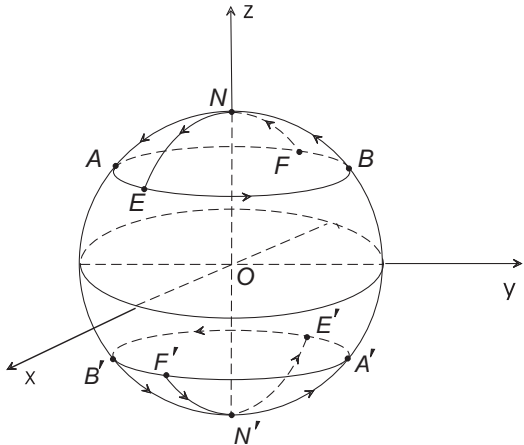}
\end{center}
\caption{The cyclic pathes in the experiments. When $\left
|\protect\psi%
_{+}\right\rangle$ loops along the path $N$-$A$-$B$-$N$, and $\left |\protect%
\psi_{-}\right\rangle$ loops along path $N^{\prime}$-$A^{\prime}$-$B^{\prime}
$-$N^{\prime}$, the phases gained are $-\protect\pi/2$ and $\protect\pi/2$,
correspondingly. The polar angle of $A$ is arbitrary between $0$ and $%
\protect\pi/2$.}
\label{path}
\end{figure}
In the same spirit as that in Ref. \cite{ZhuWang03}, once we are able to
ensure that the total phase $\gamma $ is a generalized unconventional
geometric phase given by
\begin{equation}
\gamma =\alpha _{g}+(1+\eta )\gamma _{g},  \label{Unconventional}
\end{equation}%
this $U$-gate is an unconventional geometric quantum gate. For example,
along a cyclic evolution path $A$-$B$-$N$-$A$ on the Bloch sphere in Fig.\ref%
{path}, it is found that $\gamma _{d}=-\pi /2-\gamma _{g}$, and thus the
unconventional geometric phase ($-\pi /2$) will be accumulated in the
qubit-state. The corresponding $U$-gate is just an unconventional GQC gate.
In the experiment, the two single-qubit gates to be chosen are $U_1=\left(%
\begin{array}{cc}
-i & 0 \\
0 & i%
\end{array}%
\right)$, ($\gamma=-\pi/2$, $\chi=0$, $\phi=0$) and $U_2=-i/\sqrt{2}\left(%
\begin{array}{cc}
1 & 1 \\
1 & -1%
\end{array}%
\right)$, ($\gamma=-\pi/2$, $\chi=\pi/4$, $\phi=0$).  As is well known, to
achieve a set of universal quantum gates, in addition to the above two
noncommutative single-qubit gates, we need also to construct one nontrivial
two-qubit gate based on unconventional geometric phase shifts. In the
present work, a nontrivial two-qubit gate is obtained when the loop is
controlled by another qubit; for example the controlled loop $N$-$A$-$B$-$N$
leads to a controlled unconventional GQC gate:
\begin{equation}
U_{c}=\left(
\begin{array}{cccc}
-i & 0 & 0 & 0 \\
0 & i & 0 & 0 \\
0 & 0 & 1 & 0 \\
0 & 0 & 0 & 1%
\end{array}%
\right)  \label{two}
\end{equation}

We now demonstrate how to achieve the above unconventional GQC
gates using NMR interferometer, noting that the NMR has been a
mature technology to simply illustrate certain quantum information
processing methods in recent years and a number of important
experiments like the demonstration of the Shor algorithm have been
reported\cite{RMP2004}. In our NMR experiment, the two -qubit NMR
system is 0.5ml, 20mmol sample of carbon-13 labelled chloroform
(CHCl$_{3}$) dissolved in d$_{6}$ acetone~\cite{Du2}. Qubit $a$ is
identified as the carbon-13 nucleus and qubit $b$ as the hydrogen.
The Hamiltonian of the system is written as:

\begin{equation}
H=\omega _{a}I_{z}^{a}+\omega _{b}I_{z}^{b}+2\pi JI_{z}^{a}I_{z}^{b},
\end{equation}
where the first two terms characterize the free procession of spin
carbon-13 and hydrogen about the externally applied, strong static
$B_{0}$ with frequencies $\omega _{a}/2\pi \simeq 100MHZ$ and \
$\omega _{b}/2\pi \simeq 400MHZ$, and $I_{z}^{a}$ and $I_{z}^{b}$\
are the $z$ components of the angular momentum operators for qubit
$a$ and qubit $b$ ($I_{z}^{a}=1/2\sigma
_{z}^{a},I_{z}^{b}=1/2\sigma _{z}^{b})$. The third term
characterizes a scalar spin-spin coupling of the two spins with
$J=214.5HZ$. The spin-spin relaxation time are 0.35s for carbon
and 3.3s for proton, respectively.

Initially the two qubits are in thermal equilibrium with the environment and
their state is described by the density operator $\rho _{th}\varpropto
I_{z}^{a}+4I_{z}^{b}$. We use the ''spatial averaging'' technique to prepare
the effective pure state  $|00\rangle$, or in the density operator form: $%
\frac{1}{2}(1+2I_{z}^{a})\otimes \frac{1}{2}(1+2I_{z}^{b})$. The
NMR pulse sequence to generate this state is: $R_{x}^{b}(\frac{\pi
}{3}
)-G_{z}-R_{x}^{b}(\frac{\pi }{4})-\frac{1}{2J}-R_{-y}^{b}(\frac{\pi }{4}%
)-G_{z},$ where $R_{x}^{b}(\alpha )=e^{-i\alpha I_{x}}$ denotes a hard pulse
applied on qubit $b$ to make it rotate around the $x$ axis by angle $\alpha $
($R_{-x}^{b}(\alpha )=R_{x}^{b}(-\alpha )$). $G_{z}$ indicates a $z$%
-gradient which destroys all coherences ($x$ and $y$ magnetizations) and
retains only longitudinal magnetization ($z$ magnetization component), and $%
\frac{1}{2J}$ represents a time interval  during which only the
third term of the Hamiltonian evolves.

In the beginning, we measure the unconventional geometric phase
when the state  evolves along the loop we have designed. The
quantum  network utilized and a brief measurement description are
presented  in Fig.2. Qubit $a$ is the auxiliary qubit to help
observe the phase of qubit  $b$ acquired when undergoing a cyclic
evolution path in Fig.1. The Hamiltonian for spin $b $ in
rotational framework of rotation speed
$\omega^{\prime}_b=\omega_b-\pi J $ is given by
\begin{equation}
H_b=(\omega_b-\omega^{\prime}_b\pm \pi J)I_{z}^{b}
\end{equation}%
Note here that $H_b$ is dependent on the state of qubit $a$ through the term
$\pm J$. Explicitly, it is $2\pi JI_{z}^{b}$ if the state of qubit $a$ is $%
\left
|\psi_{a}\right\rangle=\left|\uparrow\right\rangle$, and it is zero
if the state of qubit $a$ is $\left
|\psi_{a}\right\rangle=\left|\downarrow%
\right\rangle$.

\begin{figure}[tbp]
\begin{center}
\includegraphics[width=0.85\columnwidth]{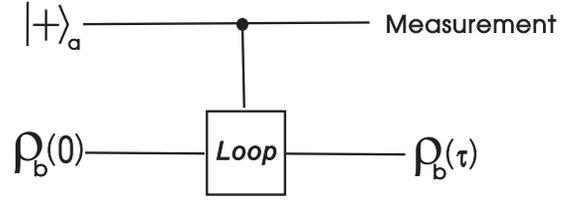}
\end{center}
\caption{The quantum network to measure the unconventional geometrical
phase, which is in fact a two-qubit gate. The top horizontal line represents
an auxiliary spin half particle, or an auxiliary qubit, labeled as qubit $a$%
. The bottom line represents a qubit labeled as $b$, in state $\protect\rho%
_{b}$ which undergoes a cyclic evolution induced by a unitary operation $U(%
\protect\tau)$. We choose our reference basis, for qubits $a$ and $b$, to be
states $\left|\uparrow\right\rangle$ and $\left|\downarrow\right\rangle$. In
this basis $\left|\pm\right\rangle=1/\protect\sqrt{2}\left(\left|\uparrow%
\right\rangle\pm\left|\downarrow\right\rangle\right)$, thus the initial
state of auxiliary qubit is $\left|+\right\rangle_{a}$. The phase gained by
qubit $b$ after the evolution operation $U(\protect\tau)$ is detected by a
phase sensitive detector.}
\label{data}
\end{figure}

The NMR RF pulse sequence can be expressed as
\begin{equation}
R_{y}^b(-\frac{\pi}{2})-\tau_1-R_{y}^b(\frac{\pi}{2})-\tau_2-R_{y}^b(-\frac{%
\pi}{2})-\tau_1-R_{y}^b(\frac{\pi}{2}).
\end{equation}
We interpret the process in detail as follows. Supposing that the starting
point is $N$, rotating firstly qubit $b$ around the axis $y$ with the angle
$\frac{\pi}{2}$ transforms the Hamiltonian of qubit $b$ to $2\pi JI_{x}^{b}$
if it is not zero. This operation is denoted by  $R_{y}^b(-\frac{\pi}{2})$.
Note that  $\left|\omega_b-\omega_a\right|$ is much larger than $J$, hence
the  state of qubit $a$ is (almost) unaffected by any operation on qubit $b$
in the whole process.  The interaction Hamiltonian will create an evolution
path on  the geodesic curve $N$-$A$ during the interval $\tau_1=\frac{\theta%
}{2J}$. The next step is to  rotate qubit $b$ around the $y$ axis with $%
\frac{\pi}{2}$ to let  the Hamiltonian return to the original one.  During
the interval $\tau_2=\frac{\pi}{2J}$, the state evolves freely through the
path $A$-$B$.  Then repeat the same operations as mentioned above  $R_{y}^2(%
\frac{\pi}{2})-\tau_1-R_{-y}^2(\frac{\pi}{2})$,  qubit $b$ will return to
the starting point with a phase $\gamma$ being accumulated. While the phase
is zero when qubit $a$ is $\left
|\psi_{a}\right\rangle=\left|\downarrow%
\right\rangle$. As a results, the auxiliary qubit acquires an internal phase
factor $e^{i\gamma}$, i.e., $1/\sqrt{2}\left(\left|\uparrow\right\rangle+%
\left|\downarrow\right\rangle\right)\mapsto 1/\sqrt{2}\left(e^{i\gamma}|%
\uparrow\rangle+|\downarrow\rangle\right)$. Exploiting a phase sensitive
detector on qubit $a$, we can observe/determine this phase $\gamma$.

\begin{figure}[tbp]
\begin{center}
\includegraphics[width=0.8\columnwidth]{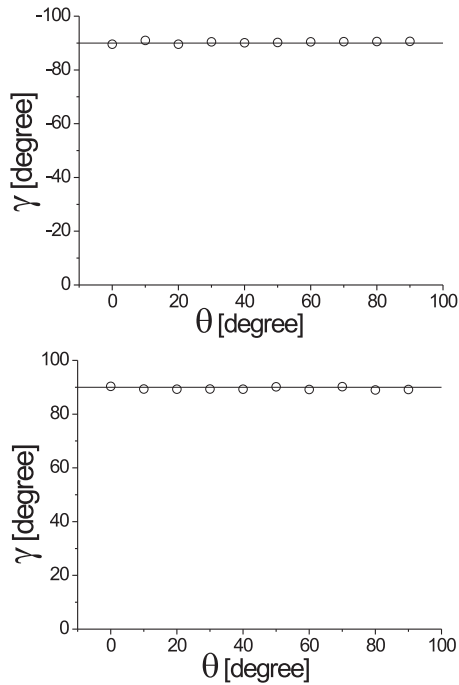}
\end{center}
\caption{The measured unconventional geometric phases in our NMR experiment.
The two reference states evolve respectively along the path $N$-$A$-$B$-$N$
(up) and $N^{\prime}$-$A^{\prime}$-$B^{\prime}$-$N^{\prime}$ (down) [see
Fig.1]. We set the polar angle $\protect\theta$ of $A$ to be $n\protect\pi/18
$ $(n=0,1,2\cdots,9)$ and $m\protect\pi/18$ $(m=9,10,11\cdots,18)$ for $%
A^{\prime}$. The measured results are in excellent agreement with Eq.(2)
with $\protect\alpha_g=-\protect\pi/2$ and $\protect\eta=-1$.}
\label{data}
\end{figure}

In the experiment, the rotation operations on qubit $b$ were performed by
hard pulses. The phases accumulated via the two designated paths: $N$-$A$-$B$%
-$N$ and $N^{\prime }$-$A^{\prime }$-$B^{\prime }$-$N^{\prime }$ were
measured. In the loop $N$-$A$-$B$-$N$, the polar angle of $A$ is set as $%
\theta =n\pi /18$ $(n=0,1,2\cdots ,9)$, and in the loop $N^{\prime }$-$%
A^{\prime }$-$B^{\prime }$-$N^{\prime }$, the polar angle of
$A^{\prime }$ as $\theta =m\pi /18$ $(m=9,10,11\cdots ,18)$. The
results are depicted in Fig.3, which are in excellent agreement
with Eq.(2) (with $\alpha _{g}=-\pi /2$ and $\eta =-1$).
Therefore, this evolution loop can be used to construct the set of
unconventional geometrical phase gates $U_{1}$, $U_{2}$, and
$U_{c} $, which will be demonstrated in the experiment detailed
below. In addition,
we adopt 
the averaged gate fidelity, defined as
\begin{equation}
\mathcal{F}=\overline{\langle \Psi _{in}|\hat{U}^{\dag }\rho _{out}\hat{U}%
|\Psi _{in}\rangle },
\end{equation}%
to measure the precision of the experimentally implemented gates with
respect to an ideal one, where the over-line denotes the average over all
possible input states $|\Psi _{in}\rangle $, and $\hat{U}$ is the unitary
operator corresponding to the ideal gate.

In our experiment, two single-qubit gates were implemented by performing the
following procedures. In consideration of the symmetry of the two qubits ,
both gates were implemented on spin (qubit) $a$, with spin (qubit) $b$ being
decoupled by spin echoes. In both processes to realize the single-qubit
gates, the irradiation frequency of the carbon channel was set as $\varpi
_{a}-4\pi J$ and $\varpi _{b}$ for the hydrogen channel, so the Hamiltonian
in this rotating frame reads $H_{a}=4\pi JI_{z}^{a}$.

(i) By applying the NMR pulse sequence $R_{x}^{a}(\frac{\pi }{4})-\frac{1}{8J%
}-R_{x}^{b}(\pi )-\frac{1}{8J}-R_{-x}^{b}(\pi )-R_{x}^{a}(\frac{\pi }{4})$,
in which the $\pi$ pulses on spin $b$ are used to cancel the coupling J of
the two qubits. Under the pulse sequence the starting state $%
\left|0\right\rangle$ cycles along path $N$-$A$-$B$-$N$ with the state $%
\left|1\right\rangle$ being meanwhile along the path $N^{\prime}$-$A^{\prime}
$-$B^{\prime}$-$N^{\prime}$. Since $\gamma=-\pi/2$,$\chi=\phi=0$, we achieve
the gate operation $U_1$ from Eq.(1). The average fidelity of one qubit gate
may be evaluated by simply averaging the fidelities of six axial pure states
on the Bloch Sphere \cite{mark}: $|0\rangle$,$|1\rangle$,$%
|0\rangle\pm|1\rangle$,$|0\rangle\pm i|1\rangle$. In the experiment, by
quantum state tomography, the fidelities of the six pure states were
measured as $0.999$, $0.999$, $0.982$, $0.990$, $0.969$ and $0.983$, thus
the average fidelity is $0.985$.

(ii) By applying another NMR pulse sequence $\frac{1}{8J}$-$R_{x}^{2}(\pi )-%
\frac{1}{8J}-R_{-x}^{2}(\pi )-R_{y}^{a}(\frac{\pi }{2})$, the state $%
\left|\psi_{+}\right\rangle=\cos\frac{\pi}{8}\left|\uparrow\right\rangle+\sin%
\frac{\pi}{8}\left|\downarrow\right\rangle$ cycles along the path $E$-$F$-$N$%
-$E$ while $\left|\psi_{-}\right\rangle=\sin\frac{\pi}{8}\left
|\uparrow%
\right\rangle-\cos\frac{\pi}{8}\left |\downarrow\right\rangle$ along the
mirror loop $E^{\prime}$-$F^{\prime}$-$N^{\prime}$-$E^{\prime}$ on south
hemisphere. As a result, the logic gate $U_2$ is realized, with the fidelity
$0.975$. \


(iii) We now turn to address how to realize a nontrivial two-qubit
gate experimentally. Figure 2 shows a schematic network of a
two-qubit gate. If qubit $a$ is in the state
$\left|\uparrow\right\rangle$, an evolution
path of $N$-$A$-$B$-$N$ (or $N^{\prime}$-$A^{\prime}$-$B^{\prime}$-$%
N^{\prime}$) on the Bloch sphere is produced for qubit $b$; if qubit $a$ is
in the state $\left|\downarrow\right\rangle$, nothing happens to qubit $b$.
This is equivalent to say that the time evolution operator satisfies the
relation: $U(\tau)\left|\psi_{\pm}\right\rangle=e^{\mp
i\pi/2}\left|\psi_{\pm}\right\rangle$ if qubit $a$ is up, while $U(\tau)=1$
if qubit $a$ is down. Here $\left|\psi_{\pm}\right\rangle$ correspond to
point $N$ and $N^{\prime}$ respectively in the Bloch sphere, with $\mp\pi/2$
are the unconventional geometrical phases acquired respectively
Therefore the two-qubit logical gate $U_c$ is obtained.

This is a nontrivial conditional phase gate(two-qubit)\cite{Jones, S.L.Zhu}.
To completely characterize the process, we performed a process tomography on
the two qubit gate. Using the method described in Ref. \cite{prl78}, we
realized quantum state tomography of the $16$ states: $\left\vert \psi
_{k}\right\rangle \left\vert \psi _{l}\right\rangle $ $(k,l=1,\cdots ,4)$,
where $\left\vert \psi _{1}\right\rangle =\left\vert 0\right\rangle $, $%
\left\vert \psi _{2}\right\rangle =\left\vert 1\right\rangle $, $\left\vert
\psi _{3}\right\rangle =\frac{1}{\sqrt{2}}(\left\vert 0\right\rangle
+\left\vert 1\right\rangle )$, $\left\vert \psi _{4}\right\rangle =\frac{1}{%
\sqrt{2}}(\left\vert 0\right\rangle +i\left\vert 1\right\rangle
)$. The average fidelity of the gate was found to be $0.934$,
which is higher than that of the conventional dynamical
gate.

All our measurements are conducted at room temperature and normal pressure
on a Brucker AV-400 spectrometer, and quite high fidelities of a universal
set of quantum gates are achieved in NMR systems. Essentially this is due to
the fact that such kind of gates\ are realized by using unconventional
geometric phase to fight against decoherence. Therefore, the measured
fidelities are robust. However, there are some observed small deviations.
Such residual imperfections will always remain in the experiment and may
come from imperfect pulses, quantum tomography and inhomogeneity of magnetic
field.

In conclusion, we have generalized the idea of unconventional geometric
phase gates. It has been illustrated that this generalized unconventional
GQC can be implemented in NMR systems. Indeed, we carried out the first
experiment to achieve a universal set of such kind of gates with quite high
fidelity in NMR systems. Our scheme using the generalized unconventional
geometric phase is very interesting and valuable in physical implementation
of geometric quantum computation because it can shorten the gate-operation
time while keeping the high fidelity. The present scheme may also be
feasible in other physical systems, which would stimulate significant
experimental interests.

We are grateful to Mianlai Zhou and S. L. Zhu for helpful discussions. This
work was supported by the National Fundamental Research Program (Grant No.
2001CB309300), ASTAR under Grant No. R-144-000-071-305, NSFC under
Grant Nos. 10425524 \& 10429401, the RGC grant of Hong Kong under Nos.
HKU7114/02P \&HKU7045/05P, and the URC fund of HKU.



\begin{thebibliography}{99}
\bibitem{Shor1999} P. W. Shor, SIAM Rev.\textbf{41}, 303(1999).

\bibitem{Berry1984} M. V. Berry, Proc. R. Soc. London, Ser.A \textbf{392},
45(1984).

\bibitem{Aharonov} Y. Aharonov and J. Anandan, Phys. Rev. Lett. \textbf{58},
1593 (1987).

\bibitem{Zhu2000} S. L. Zhu, Z. D. Wang, and Y. D. Zhang, Phys. Rev. B
\textbf{61}, 1142 (2000); S. L. Zhu and Z. D.Wang, Phys. Rev. Lett. \textbf{%
85}, 1076 (2000).

\bibitem{Zanardi} P. Zanardi and M. Rasetti, Phys. Lett. A \textbf{264}, 94
(1999).

\bibitem{Falci} G. Falci, R. Fazio, G. M. Palma, J. Siewert, and V. Vedral,
Nature(London) \textbf{407}, 355(2000).

\bibitem{Cirac} L. M. Duan, J. I. Cirac, and P. Zoller, Science \textbf{292}%
, 1695(2001).

\bibitem{Jones} J. A. Jones, V. Vedral, A. Ekert, and G. Castagnoli, Nature
(London) \textbf{403}, 869(2000).

\bibitem{X.B.Wang} X. B. Wang and M. Keiji, Phys. Rev. Lett. \textbf{87},
097901 (2001).

\bibitem{Zhu} S. L. Zhu and Z. D. Wang, Phys. Rev. Lett. \textbf{89},
097902(2002); Phys. Rev. A \textbf{66}, 042322 (2002).

\bibitem{S.L.Zhu} S. L. Zhu and Z. D. Wang, Phys. Rev. A \textbf{67},
022319(2003); X. D. Zhang, S. L. Zhu, L. Hu, and Z. D. Wang, Phys. Rev. A
\textbf{71}, 014302 (2005).

\bibitem{ZhuWang03} S. L. Zhu and Z. D. Wang, Phys. Rev. Lett. \textbf{91}%
,197902 (2003).


\bibitem{Leibfried} D. Leibfried, B. DeMarco, V. Meyer, D. Lucas, M.
Barrett, J. Britton, W. M. Itano, B. Jelenkovic, C. Langer, T. Rosenband,
and D. J. Wineland, Nature (London) \textbf{422}, 412 (2003).

\bibitem{ZhuWang05} S. L. Zhu, Z. D. Wang, and P. Zanardi, Phys. Rev. Lett.
\textbf{94},100502 (2005).

\bibitem{RMP2004} L. M. K. Vandersypen and I. L. Chuang, Rev. Mod. Phys.
\textbf{76}, 1037 (2004).


\bibitem{Du2} J. F. Du, P. Zou, M. Shi, L. C. Kwek, J. -W. Pan, C. H. Oh, A.
Ekert, D. K. L. Oi, and M. Ericsson, Phys. Rev. Lett. \textbf{91},
100403(2003).

\bibitem{prl78} J. F. Poyatos and J . I. Cirac, Phys. Rev. Lett. \textbf{78},
390(1997).

\bibitem{mark} Jae-Seung Lee, Phys. Lett. A \textbf{305}, 349 (2002); M D.
Bowdrey, D. K. L. Oi, A. J. Short, K. Banaszak, and J. A. Jones,
Phys. Lett. A \textbf{294}, 258 (2002).

\end{thebibliography}
\end{document}